# Adsorption of ions from aqueous solutions by ferroelectric nanoparticles

Sergei V. Kalinin[1*], Eugene A. Eliseev[2], and Anna N. Morozovska[3†]

[1]Department of Materials Science and Engineering, University of Tennessee,
Knoxville, TN, 37996, USA

[2]Frantsevich Institute for Problems in Materials Science, National Academy of Sciences of Ukraine,
3, str. Omeliana Pritsaka, 03142 Kyiv, Ukraine

[3] Institute of Physics, National Academy of Sciences of Ukraine,
46, pr. Nauky, 03028 Kyiv, Ukraine

**Abstract**

The fundamental aspect of physics of ferroelectric materials is the screening of uncompensated bound charges by the dissociative adsorption of ionic charges from the environment. The adsorption of ions can be especially strong when the ferroelectric undergoes the temperature-induced transition from the paraelectric phase to the ferroelectric state. Here we demonstrate that the adsorption of ions and free radicals by the polar surface of ferroelectric nanoparticles can be very efficient in aqueous solutions due to the strong ferro-ionic coupling in the nanoparticles. Obtained results can be useful for the elaboration of alternative methods and tools for adsorption of the cations ($Li^+$, $K^+$, $Na^+$, etc.), anions ($Cl^-$, $Br^-$, $J^-$), and/or free radicals ($CO^-$, $NH_4^+$, etc.) from aqueous solutions by the lead-free uniaxial ferroelectric nanoparticles. The results may become an alternative way for the environment-friendly laboratory-scale purification of different aqueous solutions from ionic contamination using controllable cyclic temperature variations.

**Keywords:** ferro-ionic coupling, ferroelectric nanoparticles, adsorption of ions, phase transitions, domain structure

[*] corresponding author, e-mail: sergei2@utk.edu

[†] corresponding author, e-mail: anna.n.morozovska@gmail.com

# I. INTRODUCTION

The intrinsic aspect of the physics of ferroelectric materials is the screening of the uncompensated bound charges by ionic-electronic charges from the environment [1] reinforced by the strong coupling between the ferroelectric dipoles and surface electrochemical species [2]. The adsorption of ions by the surface of ferroelectric perovskite and hafnia films in oxygen-deficient atmosphere can result in chemical switching of the spontaneous polarization [3, 4, 5, 6].

Here we show that the adsorption can be also very strong in aqueous solutions, where the cations ($Li^+$, $K^+$, $Na^+$, etc.), anions ($Cl^-$, $Br^-$, $J^-$, etc.), free radicals ($CO^-$, $NH_4^+$, etc.), or other charged contaminations can be readily captured by the polar surface of the nanoscale ferroelectric undergoing temperature-induced transition to the ferroelectric phase. The adsorption may become an alternative way for the environment-friendly cleaning/recycling of aqueous solutions from ionic contamination. The specific adsorption energies on ferroelectric surface can be the additional factor in selectivity for particular cations and anions. Particularly, the laboratory-scale purification of different aqueous solutions from ionic contamination using controllable cyclic temperature variations seems possible.

To reach the maximal effect of ions adsorption we require non-soluble in water ferroelectrics, which Curie temperature is close to the working (e.g., 20 – 40 $^0$C) temperature, and the spontaneous polarization component normal to the surface is homogeneous and as high as possible. Indeed, most oxide ferroelectrics, such as $BaTiO_3$, $PbTiO_3$, $BiFeO_3$, which have a very high melting temperature (above 1000 K) and Curie temperature (above 400 – 800 K), are regarded water non-soluble [7, 8, 9]. Chalcogenide ferroelectrics, such as $Sn_2P_2S_6$ and $CuInP_2S_6$, which have much lower Curie temperatures (~ 300 K), are also non-soluble in water; however, the partial substitution of sulfur by oxygen may appear at the nanocrystal surface due to the contact with water [10]. Only organic ferroelectrics, such as Roshelle salt and triglycine sulfate, are soluble in water [11].

Not less important are the requirements of the optimal surface to volume ratio, simple and controllable ways of synthesis, shape and sizes control, and environment-friendly chemical composition. From the shape factor perspective, these requirements are fulfilled for the inorganic lead-free ferroelectric nanoparticles of a quasi-ellipsoidal shape varying between nanodisks, nanospheres and nanoneedles. However, the polarization of nanodisks is mostly in-plane, so that their top and bottom surfaces are uncharged and cannot adsorb ions from the environment. The single-domain polarization of nanoneedles tends to align along the needle axis, so that only the small area near the needle ends can adsorb ions. The optimal situation can be realized in the nanospheres with the uniaxial single-domain polarization, where the charge adsorption takes place at the spherical surface except for the small segment in the equatorial plane. Exact tuning of the ferroelectric-paraelectric transition temperature $T_{FE}$ in the quasi-ellipsoidal nanoparticles with sizes (10 – 100) nm can be reached due to the size effects and strain engineering [12].

The nanoparticles in the paraelectric (PE) phase adsorb ions from the environment based on standard chemistry considerations. In the ferroelectric (FE) phase, the additional adsorption channel emerges when the uncompensated bound charge proportional to the spontaneous polarization appears at the nanoparticle surface (see **Fig.1(b)**). Thus, the cyclical heating (**Fig.1(b)**) and cooling (**Fig.1(c)**) of the nanoparticle ensemble above and below the ferroelectric-paraelectric transition temperature $T_{FE}$ can lead to the accumulation of free ions in the aqueous solution around the nanoparticles (see the bottom part of **Fig.1(c)**). Since the transition temperature of spherical ferroelectric nanoparticles is size-dependent for the sizes below 50 nm (see e.g., Refs. [13, 14] and review [15]), it is important to use the nanoparticle ensemble with a narrow size distribution function. The temperature cycling should be performed in the temperature range $T_{min} < T < T_{max}$, where $T_{min} < T_{FE}$ and $T_{max} > T_{FE}$. The initial state of ferroelectric nanoparticles is not important. Important is that their single-domain state is thermodynamically stable for $T_{min} < T < T_{FE}$. The single-domain nanoparticles can be initially screened by the ambient charges (not mandatory adsorbed from the water) and become paraelectric under heating for $T_{FE} < T < T_{max}$. Due to the osmatic effect, the accumulation of ions and free radicals from the solution can be at least partially irreversible when the partition, which is transparent for mobile charges and impermeable for the nanoparticles, separates the ferroelectric nanoparticles from the rest of the solution (see **Fig.1(b)**). Porous sieves with pore sizes less than the nanoparticles size can be used as the semi-transparent partition. After each cycle of heating and cooling the contaminated part solution with nanoparticles enriched by ions and radicals (below the partition in **Fig.1(c)**) should be separated from its cleaned part (above the partition in **Fig.1(c)**), re-heated to dissolve the adsorbed contamination and drained away. Of course, the speed of cleaning becomes smaller when the concentration of contamination decreases, and hypothetically the nanoparticles should become poly-domain or paraelectric in the ultra-clean water without mobile free charges. Thus, there is the minimal concentration of ions (and/or radicals) for which the proposed method still can effectively work.

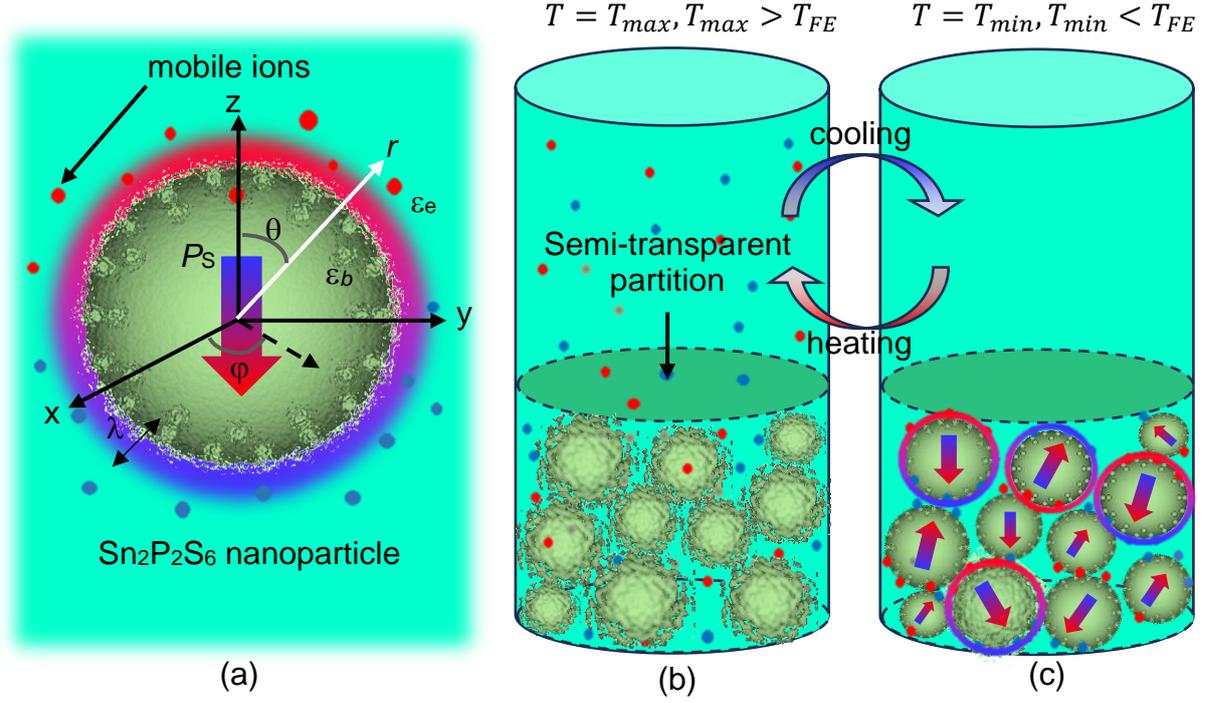

**FIGURE 1**. **(a)** The radial cross-section of the spherical ferroelectric nanoparticle covered with the shell of ionic-electronic screening charges with the effective screening length $\lambda$. The colored arrow shows the direction of the spontaneous polarization $P_S$ inside the nanoparticle. **(b)** The nanoparticles in the PE phase (i.e., at the temperature $T > T_{FE}$) and **(c)** in the FE phase (i.e., at the temperature $T < T_{FE}$) are placed the H₂O-based solution with ions and/or free radicals. The partition (e.g., the porous sieve) is transparent for mobile charges and impermeable for the nanoparticles.

We further note that the multiaxial ferroelectric nanoparticles such as (Ba,Sr)TiO$_3$ or (Bi,Sm)FeO$_3$ are less likely to be suitable for the purpose, because their spontaneous polarization rotates near the surface to create vortex-like configurations. The polar vortices do not induce any noticeable bound charge at the surface, and thus the stray field outside the nanoparticle is virtually absent [16, 17]. The absence of the uncompensated bound charge leads to the absence of the ionic charge adsorption by the nanoparticle surface. While in realistic materials the efficiency of ion adsorption is determined by the relative energies of the ion-screened phase and unscreened vortex phase, we defer detailed exploration of this scenario to future studies.

The above requirements impose at least two limitations on the properties of lead-free inorganic uniaxial ferroelectric materials which can be used for the efficient and environment-friendly ionic adsorption from aqueous solutions: their Curie temperature $T_C$ should be close to the room temperature and their spontaneous polarization $P_S$ should be relatively high near the room temperature. Among potential candidates, which synthesis routes are well-elaborated, one should consider the ferroelectric ilmenites (e.g., LiNbO$_3$ and LiTaO$_3$), chalcogenides (e.g., Sn$_2$P$_2$S$_6$), layered Van der Waals ferroelectrics (e.g., CuInP$_2$(S,Se)$_6$ family), hafnia-zirconia and other binary oxides (e.g., (Hf,Zr)O$_2$ and (Mg,Zn)O), wurtzites and fluorides [18]. However, most of these materials do not fulfil both requirements. Namely, LiNbO$_3$ and LiTaO$_3$ have a high $P_S \sim (50 – 70)$ μC/cm$^2$ at room temperature but a very high $T_C > (700 – 1200)^0$C; CuInP$_2$(S,Se)$_6$ has a very low $P_S \sim 3$ μC/cm$^2$ at room temperature and low $T_C \leq 20^0$C [19]; the optimized (Hf,Zr)O$_2$:Y has an appropriate $P_S \sim (10 – 50)$ μC/cm$^2$ at room temperature but a high $T_C > (350 – 500)^0$C [20, 21]. The chalcogenide Sn$_2$P$_2$S$_6$ (SPS) that has $P_S \sim 20$ μC/cm$^2$ at room temperature [22], $T_C = 65^0$C [23] and high background dielectric permittivity [24], seems suitable. Important that the role of size and screening effects [25] and ferro-ionic coupling [26] on the polar properties and domain structure morphology are very significant in the SPS nanoparticles. Hence, we have chosen the SPS nanoparticles, as a model system for ionic adsorption. However, we note that given the breadth of ferroelectric phenomena in materials suitable candidates can be found if the use case is established.

Below we analyze the influence of size effects and ferro-ionic coupling on the adsorption of ions from the aqueous solutions by the SPS nanoparticles and derive the general model describing ferroelectric nanoparticle behavior in aqueous solutions along the phase transitions.

## II. BASIC EQUATIONS
### A. Effective LGD free energy of the single-domain nanoparticles

In the case of the SPS nanoparticles suspension in ionic liquids or gases the Stephenson-Highland (SH) ionic adsorption can occur at the ferroelectric surface [27]. Within the SH model the dependence of the surface charge density on electric potential excess at the surface of the nanoparticle is controlled by the concentration of positive and negative surface charges. A combination of the Landau-Ginzburg-Devonshire (LGD) and SH approaches allows to derive analytical solutions describing unusual phase states in uniaxial [28, 29, 30, 31] and multiaxial [32] ferroelectric thin films and nanoparticles [33], as well as antiferroelectric thin films with electrochemical polarization switching [34, 35].

The ferro-ionic states can originate in the SPS nanoparticles from the strong nonlinear dependence of the screening charge density on the surface electric potential, also named the electrochemical overpotential [28 - 35]. The LGD free energy functional of the SPS nanoparticle, which includes the Landau-Devonshire energy ($G_{L-D}$), the Ginzburg polarization gradient energy ($G_{grad}$), the electrostatic contribution ($G_{el}$), the elastic and electrostriction ($G_{es}$) contribution, are listed in Ref.[25, 26] and reproduced in Supplemental Materials [36]. At first let us consider a spherical single-domain SPS nanoparticle. The consideration is based on the minimization of the "effective" LGD free energy,

$$F_R = \frac{1}{2}\alpha_R P^2 + \frac{1}{4}\beta P^4 + \frac{1}{6}\gamma P^6 - PE. \tag{1a}$$

The renormalized coefficient $\alpha_R$ is [37]:

$$\alpha_R(T,R) = \alpha(T) + \frac{\varepsilon_0^{-1}}{\varepsilon_b + 2\varepsilon_e + (R/\lambda)}, \tag{1b}$$

where $\alpha(T) = \alpha_T(T - T_C)$, $\varepsilon_0$ is a universal dielectric constant, $\varepsilon_b$ is a background permittivity [38], $R$ is the radius of the nanoparticle, $\lambda$ is the "effective" surface screening length and $\varepsilon_e$ is the effective dielectric permittivity of the H₂O-based environment (see **Fig.1(a)**). LGD coefficients $\alpha_T$, $\beta$ and $\gamma$ for the bulk SPS are listed in **Table SI** [36] (see also Refs. [39, 40, 41, 42, 43]). $E$ is the electric field component colinear with the polarization.

Note that the value of the particle radius $R$ determining its transition between the single-domain and poly-domain ferroelectric states is governed by the surface screening length $\lambda$ (see e.g. Refs. [44, 45]). If $\lambda$ is small enough (e.g., smaller than 0.1 nm) the single-domain state is stable for 10 nm and larger SPS nanoparticles at room and/or lower temperatures (see e.g., Fig.2 in Ref.[44]). If the screening length is larger than 0.1 nm, high values of the environmental dielectric permittivity $\varepsilon_e$ supports the stability of the single-domain state (see e.g., Fig.6 in Ref.[45]). Approximate analytical expressions derived in Refs.[44, 45] allow calculating the phase boundaries between the paraelectric (PE), poly-domain (PDFE) and single-domain (SDFE) ferroelectric states in dependence on $R$ and $\lambda$ for uniaxial ferroelectric nanoparticles. Small SPS nanoparticles, which volume is less than several so-called correlation volumes, can be in the PE phase or in the SDFE state at fixed temperature depending on their size, $\lambda$ and $\varepsilon_e$. Larger SPS nanoparticles, which volume exceeds several correlation volumes, can be in the PE phase, in the PDFE or in the SDFE state at fixed temperature depending on their size $R$, $\lambda$ and $\varepsilon_e$. Large enough nanoparticles ($R \geq 50$ nm) are mostly in the SDFE state at temperatures below (25 – 50)°C for $0 < \lambda \leq 0.1$ Å and $\varepsilon_e \cong 81$ corresponding to aqueous solutions. In what follows we check the phase state of the SPS nanoparticle analytically and numerically, with a special attention to the SDFE state stability, because the ionic adsorption is the most efficient in the state. The ionic adsorption by the nanoparticle surface in the PDFE state is much less efficient due to the significant decrease of the bare (i.e., unscreened) depolarization fields outside the poly-domain particle, which would lead to essential slowing down of the proposed cycling process and thus to the degradation of the device application over time. It also should be noted that we neglected the ions diffusion into the ferroelectric depth.

### B. Estimation of the ionic screening length and adsorbed charge

Within the SH model the dependence of the surface charge density $\sigma_S[\phi]$ on the electric potential excess $\delta\phi$ at the surface of the nanoparticle is controlled by the coverages $\theta_i[\phi]$ of positive and negative surface charges (e.g., ions) of type $i$ in a self-consistent manner. Corresponding Langmuir adsorption isotherm is [46]:

$$\sigma_S[\phi] = \sum_i \frac{eZ_i \theta_i[\phi]}{A_i} \cong \sum_i \frac{eZ_i}{A_i} \left(1 + a_i \exp\left[\frac{\Delta G_i + eZ_i \delta\phi}{k_B T}\right]\right)^{-1}, \qquad (2)$$

where $e$ is the electron charge, $Z_i$ is the ionization number of the adsorbed ions, $a_i$ is the dimensionless chemical activity of the ions in the environment, $T$ is the absolute temperature, $A_i$ is the area per surface site for the adsorbed ion, $\Delta G_i$ are the formation energies of the surface charges (e.g., ions and/or electrons) at normal conditions. The subscript $i = 1, 2$. While a gaseous environment with a given oxygen partial pressure is considered in Refs. [2-5, 27], one can use the same expression (2) for an aqueous environment with dissolved ions or/and radicals by converting partial pressure to the chemical activity $a_i$.

The linearization of Eq.(2) for small overpotentials (i.e., for $\left|\frac{eZ_i\delta\phi}{k_BT}\right| < 1$) leads to the approximate expression for the effective surface charge density $\sigma_S$ and inverse screening length $\lambda$ [44]:

$$\sigma_S[\phi] \approx -\varepsilon_0\frac{\delta\phi}{\lambda}, \qquad \frac{1}{\lambda} \approx \sum_i \frac{(eZ_i)^2 a_i \exp\left[\frac{\Delta G_i}{k_BT}\right]}{\varepsilon_0 k_B T A_i \left(a_i + \exp\left[\frac{\Delta G_i}{k_BT}\right]\right)^2}. \qquad (3)$$

Hereinafter we use equal values of the surface charge formation energies, $\Delta G_1 = \Delta G_2 \cong (0.03 - 0.3)$ eV, opposite ionization numbers, $Z_1 = -Z_2 = 1$, put $A_1 = A_2 \cong 0.16$ nm$^2$ [5, 27] and vary $a_i$ from 0 to 1. Considering the quadratic dependence of the screening length $\lambda$ on $Z_i$, namely $\lambda \sim \frac{A_i}{(eZ_i)^2 a_i}$, obtained results are the same for $Z_1 = -Z_2 = -1$. Approximate expression (3) causes a simple scaling of the obtained dependences with the changes in ionization numbers $Z_i$ (e.g., for $Z_1 = -Z_2 = 2$ and $Z_1 = -Z_2 = -2$), ions activities $a_i$ and areas per surface site $A_i$. Higher Z and/or smaller $A_i$ decrease $\lambda$ and increase the screening degree.

The range (0.01 – 0.1)eV of the surface charge formation energies are consistent with the literature data [5, 27] for adsorption of oxygen ions (or vacancies) by the PbTiO$_3$ surface. We would like to underline that the formation energies $\Delta G_i$, which correspond to the adsorption of a concrete charge species (e.g., a single ion or radical) by a ferroelectric surface, should be obtained from the first-principles studies or experimentally (see e.g., an early review [47] and more recent works [48, 49, 50, 51, 52] devoted to the experimental and/or theoretical studies of the adsorbed ions/molecules onto ferroelectric surfaces). In particular, Wang et. al. [49] revealed experimentally that typical molecular adsorption energies for the BaTiO$_3$ surface vary from 0.1 eV to 0.2 eV (for a water molecule at the ferroelectric surface), meanwhile dissociative adsorption energies are much higher and can reach 1.0 eV for the water molecule. The differences in desorption activation energies (or heats of adsorption) between positively and negatively poled ferroelectric surfaces are relatively small, e.g., Garrity et. al. estimated them to be 110 K (i.e., 0.009 eV) for the 2-propanol adsorbed at positive (485 K, 0.042 eV) and negative (375 K, 0.03 eV) polar cuts of LiNbO$_3$ [48]. Bonnell et.al. [53] have characterized the interaction of a series of small molecules on the LiNbO$_3$, BaTiO$_3$ and (Pb,Zr)TiO$_3$ surfaces and revealed that the polar molecules of water and methanol adsorb more strongly on positively poled surfaces, but that the differences are small, only a few kJ mol$^{-1}$ (that is 0.02 – 0.04 eV per molecule). Since SPS nanoparticles have a significantly smaller spontaneous polarization at room temperature (about 15 μC/cm² or smaller) in comparison with LiNbO$_3$ (about 75 μC/cm²), BaTiO$_3$ (about 25 μC/cm²) and Pb$_x$Zr$_{1-x}$TiO$_3$ (about 50 – 75 μC/cm² in dependence on the Pb content x for x>0.5), associated adsorption energies $\Delta G_i$ could be significantly smaller (due to the smaller polarization energy of the unscreened bare surface). Neglecting the differences between the adsorption energies at positively and negatively polarized surfaces, which is rough approximation, we assume that the range $\Delta G_i \cong (0.03 - 0.3)$ eV is physically reasonable.

Since the system tries to minimize the overpotential value, the inequality $\left|\frac{eZ_i\delta\phi}{k_BT}\right| < 1$ is valid in the broad range of ion formation energies near the room temperature, and thus the validity range of the approximate Eq.(3) is rather broad too.

The dependence of λ on the temperature $T$ and the activity of ions $a_i$ calculated at $\Delta G_i = 0.04$ eV is shown in **Fig. 2(a)**. It is seen that λ very slightly depends on $T$ in the temperature range (0 – 100)°C and strongly decreases (below 0.1 nm) with the increase in $a_i$ above $5 \cdot 10^{-3}$. The dependence of λ on $a_i$ and $\Delta G_i$ calculated at room temperature is shown in **Fig. 2(b)**. It is seen that λ very strongly increases with the increase in $\Delta G_i$ above 0.06 eV.

When λ is very small (e.g., significantly less than 0.1 nm), the environment provides a very good screening of the spontaneous polarization and thus prevents the domain formation [54, 55]. Thus, the assumption of the single-domain state in the SPS nanoparticle can be self-consistent for $a_i > 3 \cdot 10^{-2}$ and $\Delta G_i < 0.06$ eV. To account for the possible domain formation, which can appear for λ ≥0.1 nm, one should use the finite element modeling (FEM) based on the LGD free energy and Poisson equation listed in Suppl. Mat. [36].

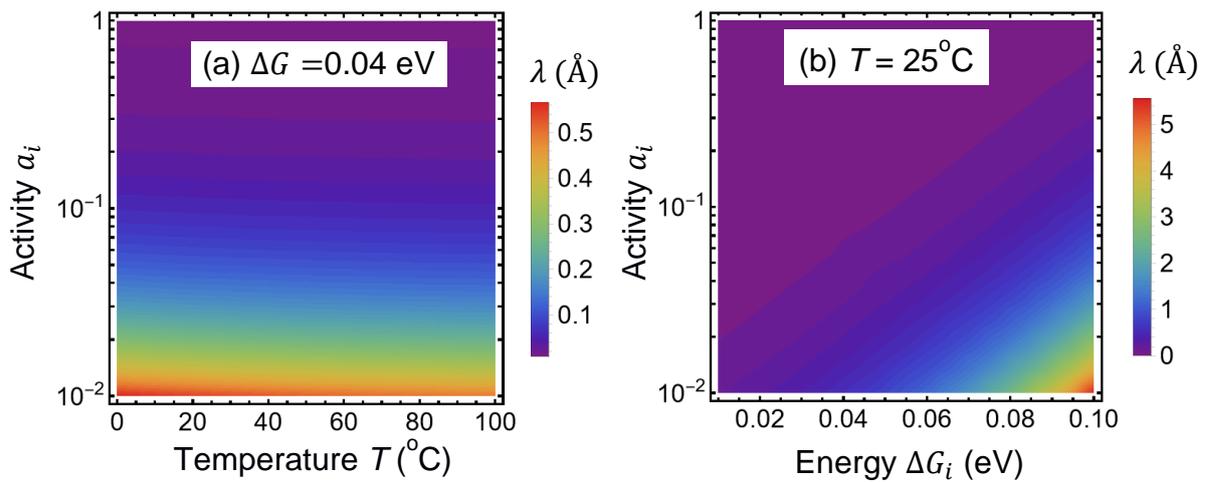

**FIGURE 2**. The dependence of λ on the activity of ions $a_i$, temperature $T$ **(a)** and surface charge formation energy $\Delta G_i$ **(b)** calculated for $Z_1 = -Z_2 = 1$ and $A_1 = A_2 \cong 0.16$ nm². The energy $\Delta G_i = 0.04$ eV for the plot **(a)** and $T = 25$°C for the plot **(b).**

To calculate the adsorbed surface charge from Eq.(3) the electric potential excess $\delta\phi$ at the surface of the nanoparticle should be determined. The rigorous calculation of $\delta\phi$ is a rather complicated self-consistent problem allowing for the ion dynamics in the aqueous solution. In this section, where we consider the single-domain nanoparticles, we will use the simplest estimation valid for small $\lambda \ll 0.1$ nm. In the case of small screening length, the angular distribution of the surface charge, $\bar{\sigma}_S(\varphi, \theta)$, absorbed by the single-domain nanoparticle is almost equal to its bound charge $P_S\cos\theta$, being proportional to the quasi-homogeneous spontaneous polarization $P_S$. Note that the well-screened nanoparticles do not interact with each other. Thus, the magnitudes of the average spontaneous polarization and the absorbed charge can be estimated as:

$$\bar{P}_S = \frac{-\beta + \sqrt{\beta^2 - 4\alpha_R \gamma}}{2\gamma}, \tag{4a}$$

$$|\bar{\sigma}_S| \cong \frac{1}{2\pi} \int_0^{2\pi} d\varphi \int_0^{\pi/2} \bar{P}_S(T, R) \cos\theta \sin\theta d\theta = P_S. \tag{4b}$$

The averaging in Eq.(4b) is performed over one of the semi-spheres, where the bound charge $P_S\cos\theta$ does not change its sign (see **Fig. 1(a)**). The sum of the positive and negative bound charges is $2P_S$. Due to the Gauss theorem, the equality $|\bar{\sigma}_S| = P_S$ is valid for an arbitrary semi-ellipsoidal surface. Since $\alpha_R$ depends on the $R$, $T$ and $\lambda$ in accordance with Eq.(1b), and $\lambda$ depends on the activity $a_i$ in accordance with Eq.(3), the value of $\bar{\sigma}_S$ is determined by the $R$, $T$ and $a_i$.

### III. RESULTS AND DISCUSSION
### A. The adsorption of ions by single-domain nanoparticles

Dependences of the 2D-density $\bar{\sigma}_S$ of adsorbed charges on the activity of ions $a_i$, temperature $T$, particle radius $R$ and ions formation energy $\Delta G_i$ are shown in **Fig. 3**. The panels **(a)-(d)** are calculated for the single-domain nanoparticle using the LGD free energy (1) with coefficients taken from **Table SI** [36], $Z_1 = -Z_2 = 1$, $A_1 = A_2 \cong 0.16$ nm$^2$, $\varepsilon_b = 41$ and $\varepsilon_e = 81$ (water). In fact, the color maps in **Fig. 3** are phase diagrams plotted in different coordinates, where the dark-violet regions with $\bar{\sigma}_S = 0$ corresponds to the PE phase and the colored regions with $\bar{\sigma}_S > 0$ corresponds to the FE phase of the nanoparticle. Since $\beta > 0$ for the bulk SPS, the PE-FE phase boundary corresponds to the second order phase transition.

From **Fig. 3(a)** calculated for $R = 50$ nm and $\Delta G_i = 0.04$ eV, the adsorbed charge $\bar{\sigma}_S$ occurs at $a_i \geq 10^{-2}$ and $T = 0$°C. The magnitude of $\bar{\sigma}_S$ monotonically increases with the increase in $n_i$ and decreases with the increase in $T$. The magnitude of $\bar{\sigma}_S$ quickly saturates in the "deep" FE phase (far from the PE-FE phase boundary) and overcomes 15 µC/cm$^2$ in the vicinity of the room temperature.

From **Fig. 3(b)** calculated for $T = 25$°C and $\Delta G_i = 0.04$ eV, the adsorbed charge occurs at $a_i = \geq 10^{-2}$ and $R = 100$ nm. The magnitude of $\bar{\sigma}_S$ monotonically increases and saturates (the saturation value ~15 µC/cm²) with the increase in $a_i$ and strongly decreases with the decrease in $R$. The decrease of $\bar{\sigma}_S$ becomes very steep when $R$ decreases below 20 nm. The increase in $a_i$ makes the size-induced decrease of $\bar{\sigma}_S$ less pronounced due to the depolarization field decrease with decrease in $\lambda$ (see Eqs.(1b) and (3)).

From **Fig. 3(c)** calculated for $R = 50$ nm and $T = 25^0$C, the adsorbed charge occurs at $a_i > 10^{-2}$ and $\Delta G_i = 0.05$ eV. It monotonically increases with the increase in $a_i$ and decreases strongly with the increase in $\Delta G_i$. The magnitude of $\bar{\sigma}_S$ quickly saturates to the value ~ 16 µC/cm² in the FE phase (outside the immediate vicinity of the PE-FE phase boundary).

From **Fig. 3(d)** calculated for $\Delta G_i = 0.04$ eV and $a_i = 0.1$, the adsorbed charge occurs at $R \geq 6$ nm; its magnitude monotonically increases with the increase in $R$ and decreases with the increase in $T$. Above $T_C \approx 63^0$C the charge adsorption is impossible. The magnitude of $\bar{\sigma}_S$ relatively slowly saturates to the value ~17 µC/cm² in the deep FE phase.

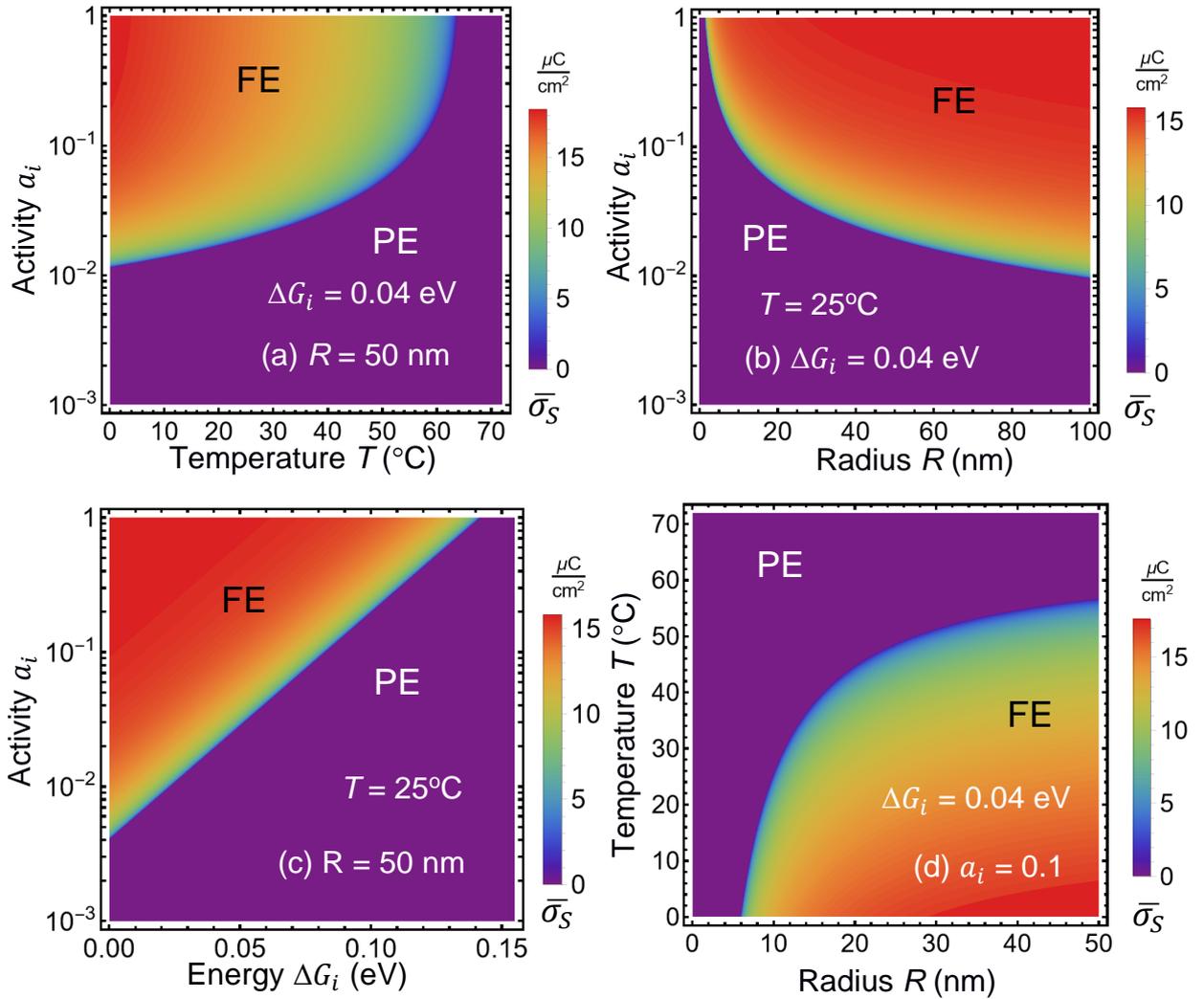

**FIGURE 3.** Dependences of the adsorbed charge density $\bar{\sigma}_S$ on the activity of ions $a_i$, temperature $T$, particle radius $R$ and ions formation energy $\Delta G_i$. The phase diagrams are calculated for SPS nanoparticles using the LGD

free energy (1) with the coefficients taken from **Table SI**, $Z_1 = -Z_2 = 1$, $A_1 = A_2 \cong 0.16$ nm², $\varepsilon_b = 41$, $\varepsilon_e = 81$. Parameters $R = 50$ nm and $\Delta G_i = 0.04$ eV for the plot **(a)**; $T = 25^0$C and $\Delta G_i = 0.04$ eV for the plot **(b)**; $R = 50$ nm and $T = 25^0$C for the plot **(c)**; $\Delta G_i = 0.04$ eV and $a_i = 0.1$ for the plot **(d)**.

## B. The adsorption of ions by poly-domain nanoparticles

Results analyzed in the previous subsection is valid only for the PE phase and "deep" single-domain FE state of the nanoparticle. However, when the activity of ions decreases below some critical value, which depends on the nanoparticle size $R$ and temperature $T$, the screening of the spontaneous polarization becomes insufficient to support the single-domain FE state in the nanoparticle. The domain emerges, and their sign-alternating bound charge also can lead to the adsorption of ions and free radicals from the aqueous solution.

To illustrate this, **Figs. 4(a)-(d)** show typical evolution of the spontaneous polarization inside the SPS nanoparticle and electric overpotential in the aqueous environment ($\varepsilon_M = 81$) with different activity of mobile ions $a_i = (10^{-3} - 1)$. The distributions of polarization and potential are calculated by FEM using the LGD free energy (S.1) [36] with the coefficients taken from **Table SI**. For $R = 25$ nm and $T = 25^0$C fine and faint ferroelectric labyrinthine domains appear inside the nanoparticle core at $a_i \geq 0.01$, while its periphery in the equatorial cross-section is almost paraelectric (see **Fig. 4(a)**). The labyrinthine contrast and area increase significantly with increase in $a_i$ (see **Fig. 4(b)** for $a_i = 0.014$). The labyrinthine domains transform into the much more regular domain structure coexisting with the single-domain state with further increase in $a_i$ (see **Fig. 4(c)** for $a_i = 0.15$). Eventually, the formation of the domain walls becomes unstable and the screening-assisted transition to the single-domain state occurs (see **Fig. 4(d)** for $a_i = 1$).

The evolution of the domain structure, which happens with the increase in $a_i$, is explained by the decrease in the screening length λ shown in **Fig. 4(e).** The decrease in λ below 0.01 nm happens at $a_i \geq 0.05$. As it follows from **Fig. 4(e)**, the adsorption can also be made ion-selective by the choice of appropriate $\Delta G_i$, and so it may be possible to remove some sorts of ions are first.

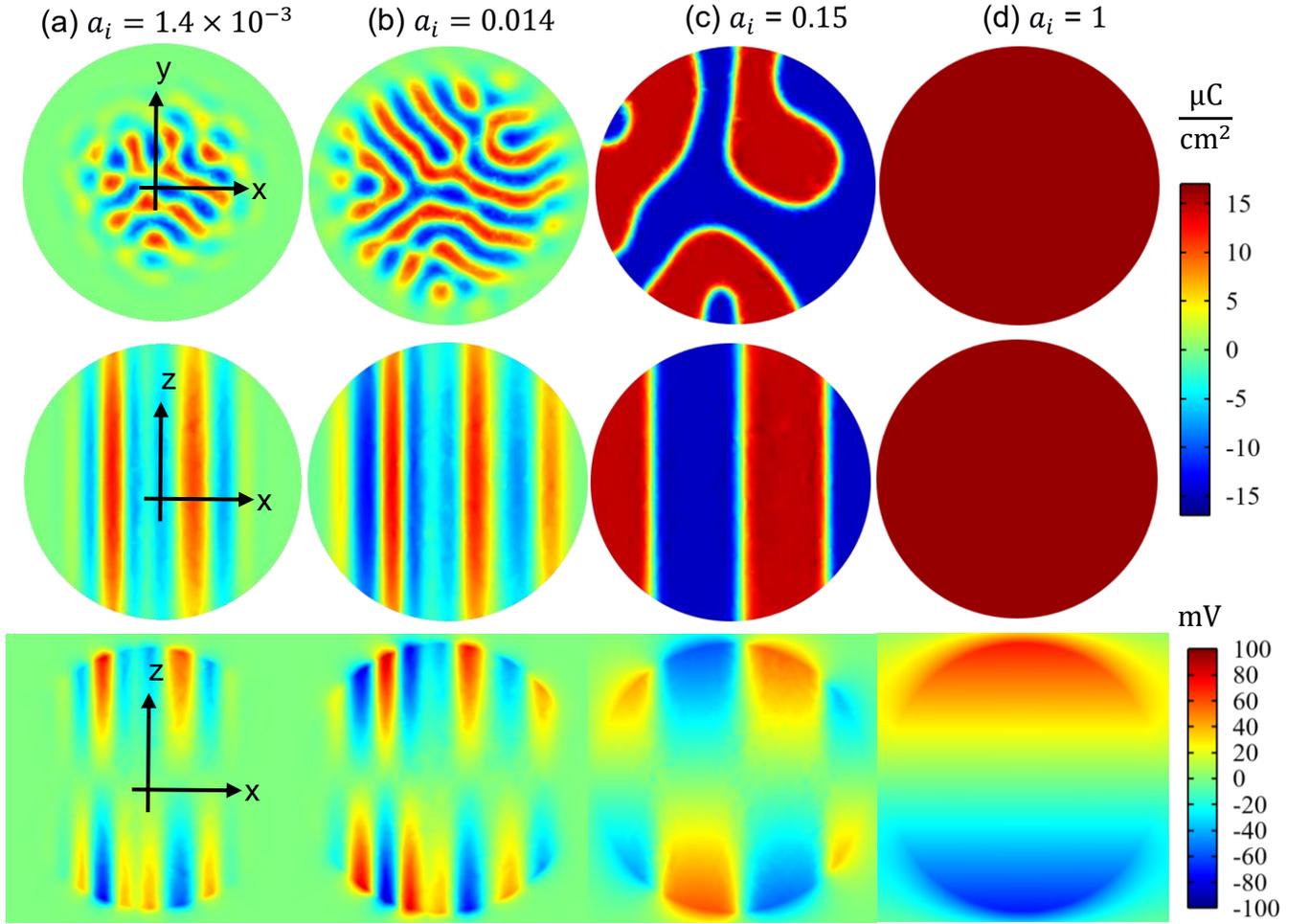

**FIGURE 4.** Typical distributions of the spontaneous polarization in the equatorial (the first row) and polar (the second row) cross-sections, and the electrostatic potential in the polar cross-section (the third row) calculated in the SPS nanoparticle covered with the adsorbed ionic-electronic charges with the activity $a_i = 10^{16}$ m$^{-2}$ **(a)**, $10^{17}$ m$^{-2}$ **(b)**, 0.15 **(c)** and 1 **(d)**. Dependence of the screening length λ **(e)** and the adsorbed charge density $\overline{\sigma_S}$ **(f)** on the activity of ions $a_i$ calculated for the formation energy $\Delta G_i$ =0.03 eV (black curves), 0.04 eV (green

curves), 0.06 eV (blue curves), 0.10 eV (magenta curves), 0.20 eV (red curves) and 0.30 eV (brown curves). Dashed curves in the plot **(f)** are calculated in the single-domain approximation of the polarization, and solid curves are calculated considering the possible appearance of the domain structure in the nanoparticle; $R = 25$ nm and $T = 25^0$C. The distributions **(a)-(d)** are calculated using the LGD free energy (S.1) with the coefficients taken from **Table SI**, $Z_1 = -Z_2 = 1$, $A_1 = A_2 \cong 0.16$ nm$^2$, $\varepsilon_b = 41$, $\varepsilon_e = 81$, $R = 25$ nm, $\Delta G_i = 0.04$ eV and $T = 25^0$C.

Dependence of the adsorbed charge density $\bar{\sigma}_S$ on the activity of ions $a_i$ calculated for the formation energies $\Delta G_i$ from 0.03 eV to 0.3 eV is shown in **Fig. 4(f)**. Dashed curves in the plot, which correspond to the single-domain state of the nanoparticle, show that $\bar{\sigma}_S$ appears under the second order PE-FE phase transition at $a_i = a_{cr}(\Delta G_i)$, increases monotonically and saturates to the bulk polarization value (~17 µC/cm$^2$) with increase in $a_i$. Solid curves, which are calculated directly from Eq.(3) by the FEM, consider the possible appearance of the domain structure in the nanoparticle. These curves behave quite different from the dashed curves. The solid curves show that $\bar{\sigma}_S$ appears under the diffuse PE-FE phase transition, because the region of the poly-domain FE phase stability is located between the PE and the single-domain FE phases. It is very important that the solid curves exist in the range of much smaller $a_i$ than the dashed curves. However, the magnitude of $\bar{\sigma}_S$ in the poly-domain FE state is smaller (or even much smaller for small $a_i$) than in the single-domain phase. This result means that the ions adsorption by the nanoparticle in the poly-domain FE state is not as effective as in the case of a single-domain nanoparticle, but still can be used for the further cleaning of the aqueous solution with lower concentration of contamination ions, free radicals and/or salts.

### C. Estimates of the nanoparticle volume ratio required for the laboratory-scale purification of aqueous solutions

An estimate for the nanoparticles volume ratio for the purification of aqueous solutions is required. Assuming that the surface density of adsorbed ions cannot exceed the maximal spontaneous polarization of SPS at room temperature, namely $P_S \cong 15$ µC/cm$^2$, this corresponds to $\frac{P_S}{eN_A} \approx 1.56 \cdot 10^{-10}$ mol/cm$^2$ of adsorbed ions ($N_A \approx 6.022 \cdot 10^{23}$ mol$^{-1}$ is the Avogadro number, and we put $|Z_i| = 1$). Assuming that the water contains about (0.1 – 1) mol. % of ionic contamination (such as Li$^+$, K$^+$ or Na$^+$ ions, CO$^-$, or NH$_4^+$ radicals), to extract the contamination from a liter of water the surface area of nanoparticles, as high as $3.56 \cdot 10^8$ cm$^2$ (for 0.1 mol. % of contamination) and $3.56 \cdot 10^9$ cm$^2$ (for 1 mol. % of contamination), is required (the volume of 1 mole of water is about 18 cm$^3$ and a liter of water is about 55.56 moles). Since the acting surface to volume ratio $\frac{3}{2R}$ is about $(6 - 3) 10^5$ cm$^{-1}$ for the SPS particles with $R = (25 - 50)$ nm, the volume of the nanoparticles required for the contamination extraction from the liter ($10^3$ cm$^3$) of water changes from $(0.59 - 1.19) \cdot 10^3$ cm$^3$ (for 0.1 mol. % of contamination) to $(0.59 - 1.19) \cdot 10^4$ cm$^3$ (for 1 mol. % of contamination), considering averaging in Eq.(4b). Thus, the ratio of nanoparticles to water volumes changes from $0.6 - 1.2$ (for 0.1 mol. % of contamination) to $6 - 12$ (for 1 mol. % of contamination). The smaller the particles are, the smaller the ratio is, but one should account for $P_S$ decrease with decrease in $R$, as well as for $P_S$ decrease with the working temperature increase.

Hence the proposed purification of the aqueous solutions may be commercially attractive only for the laboratory-scale purification of highly toxic diluted contaminations in extremal conditions (e.g., in spaceships or closed systems). To reduce the cost and to increase the efficiency of purification, multiple cycling of temperature with a short period of time is required. It may seem that the above estimates can be valid for the concentrations of contamination much smaller than 0.1 mol.%, but ultra-small concentrations of ions or radicals require much longer periods of the temperature cycling due to the finite mobilities of ion in the solution and possible strong crosstalk of multipole-type long-range electric fields caused by the partially unscreened single-domain nanoparticles with the different orientation of their spontaneous polarization. It is problematic to estimate the realistic minimal concentration of ionic contamination in the solution, which could be reached by ionic adsorption at the ferroelectric surface, as well as to predict how these estimates may vary with the system parameters (sizes, temperature, amount of the single-domain nanoparticles and ions mobility). We hope that these practically important questions can be addressed in further studies.

## IV. CONCLUSIONS

We have shown that the adsorption of ions and free radicals by the polar surface of uniaxial ferroelectric nanoparticles can be very efficient in aqueous solutions due to the strong ferro-ionic coupling in the nanoparticles. In particular, the ion adsorption by the quasi-spherical single-domain SPS nanoparticles of $50 - 100$ nm size is the most efficient for the high chemical activity of ions ($\sim 10^{-2} - 10^{-1}$ and higher) at the particle surface. In this case the adsorbed charge density is approximately equal to the average spontaneous polarization of the nanoparticle.

When the chemical activity of ions decreases below some critical value, which depends on the nanoparticle size and temperature, the screening of the spontaneous polarization becomes insufficient to support the single-domain FE state in the nanoparticle. The domains emerge, and their sign-alternating bound charge also adsorbs ions and free radicals from the aqueous solution. The ions adsorption by the nanoparticle in the poly-domain FE state is not as effective as in the case of a single-domain nanoparticle, but still can be used for the further cleaning of the aqueous solution with lower concentration of ions ($\sim 10^{15} - 10^{17}$ m$^{-2}$), free radicals and/or salts.

Obtained results can be useful for the elaboration of alternative methods and tools for adsorption of cations ($Li^+$, $K^+$, $Na^+$, etc.), anions ($Cl^-$, $Br^-$, $J^-$), and/or free radicals ($CO^-$, $NH_4^+$, etc.) from the aqueous solutions by the lead-free uniaxial ferroelectric nanoparticles. In particular, the results may become an alternative way for the environment-friendly laboratory-scale purification of different aqueous solutions from ionic contamination.

**Acknowledgements.** Authors are very grateful to the Referees for Their valuable ideas, estimates and other suggestions on the work improvement. This effort (problem statement and general analysis, S.V.K.) was supported by SVK start-up funds. The work of A.N.M. and E.A.E. are funded by the National Research Foundation of Ukraine (projects "Manyfold-degenerated metastable states of spontaneous polarization in nanoferroics: theory, experiment and perspectives for digital nanoelectronics", grant N 2023.03/0132 and "Silicon-compatible ferroelectric nanocomposites for electronics and sensors", grant N 2023.03/0127). Numerical results presented in the work are obtained and visualized using a specialized software, Mathematica 14.0 [56].

**Authors' contribution.** S.V.K. generated the research idea, formulated the physical problem and made conclusions. A.N.M. formulated equations, performed analytical calculations and prepared corresponding figures. E.A.E. wrote the codes. S.V.K. and A.N.M. wrote the manuscript draft. All co-authors discussed the results and worked on the manuscript improvement.

# SUPPLEMENTARY MATERIALS
## to the manuscript
# "Adsorption of ions from aqueous solutions by ferroelectric nanoparticles"

Sergei V. Kalinin[1*], Eugene A. Eliseev[2], and Anna N. Morozovska[3†]

[1]Department of Materials Science and Engineering, University of Tennessee,
Knoxville, TN, 37996, USA

[2]Frantsevich Institute for Problems in Materials Science, National Academy of Sciences of Ukraine,
3, str. Omeliana Pritsaka, 03142 Kyiv, Ukraine

[3] Institute of Physics, National Academy of Sciences of Ukraine,
46, pr. Nauky, 03028 Kyiv, Ukraine

**Appendix A. Free energy functional and material parameters**

The ferroelectric polarization $P_3$ contains background and soft-mode contributions, where the corresponding electric displacement $\vec{D}$ has the form $\vec{D} = \varepsilon_0 \varepsilon_b \vec{E} + \vec{P}$ inside the particle, and $\vec{D} = \varepsilon_0 \varepsilon_e \vec{E}$ outside it. Here $\vec{E}$ is the electric field, $\varepsilon_b$ is an isotropic background permittivity, and $\varepsilon_e$ is a relative dielectric permittivity of the environment. The dependence of the in-plane polarization components on $\vec{E}$ is linear, $P_i = \varepsilon_0(\varepsilon_b - 1)E_i$, where i = 1, 2.

The Landau-Ginzburg-Devonshire (LGD) free energy functional has the following form:

$$G = G_{L-D} + G_{grad} + G_{el} + G_{es} + G_{flexo}. \tag{S.1a}$$

The functional includes the Landau-Devonshire "bulk" energy, $G_{L-D}$, the polarization gradient energy (Ginzburg contribution), $G_{grad}$, the electrostatic contribution $G_{el}$, the elastic and electrostriction contributions, $G_{es}$, and the flexoelectric contributions, $G_{flexo}$ [1]:

$$G_{L-D} = \int_{r<R} dx_1 dx_2 dx_3 \left(\frac{\alpha}{2}P_3^2 + \frac{\beta}{4}P_3^4 + \frac{\gamma}{6}P_3^6\right), \tag{S.1b}$$

$$G_{grad} = \int_{r<R} dx_1 dx_2 dx_3 \left(\frac{g_{11}}{2}\left(\frac{\partial P_3}{\partial x_3}\right)^2 + \frac{g_{44}}{2}\left[\left(\frac{\partial P_3}{\partial x_2}\right)^2 + \left(\frac{\partial P_3}{\partial x_1}\right)^2\right]\right), \tag{S.1c}$$

$$G_{el} = -\int_{r<R} dx_1 dx_2 dx_3 \left(P_3 E_3 + \frac{\varepsilon_0 \varepsilon_b}{2} E_i E_i\right) - \int_{r=R} d^2r \frac{\phi}{2}\sigma - \frac{\varepsilon_0 \varepsilon_e}{2}\int_{r>R} E_i E_i d^3r, \tag{S.1d}$$

$$G_{es} = \int_{r<R} dx_1 dx_2 dx_3 \left(-\frac{s_{ijkl}}{2}\sigma_{ij}\sigma_{kl} - Q_{ij3}\sigma_{ij}P_3^2\right), \tag{S.1e}$$

---

[*] corresponding author, e-mail: sergei2@utk.edu
[†] corresponding author, e-mail: anna.n.morozovska@gmail.com

$$G_{flexo} = -\int_{r<R} dx_1 dx_2 dx_3 \frac{F_{ijk3}}{2}\left(\sigma_{ij}\frac{\partial P_3}{\partial x_k} - P_3 \frac{\partial \sigma_{ij}}{\partial x_k}\right). \tag{S.1f}$$

Hereinafter $r = \sqrt{x_1^2 + x_2^2 + x_3^2}$. The coefficient α linearly depends on temperature $T$, $\alpha = \alpha_T(T - T_C)$, where $T_C$ is the Curie temperature. The LGD-coefficients γ, β and the gradient coefficients $g_{11}$, $g_{44}$ are positive and temperature independent. An isotropic approximation, $g_{44} \approx g_{55}$ in the (001) plane is used for the monoclinic SPS structure. $\sigma_{ij}$ is the stress tensor; $s_{ijkl}$ are elastic compliances; $Q_{ijkl}$ are the electrostriction tensor components, and $F_{ijkl}$ are the flexoelectric tensor components. The electric field components $E_i$ are derived from the electric potential $\phi$ as $E_i = -\partial\phi/\partial x_i$.

The evident form of the $G_{flexo}$ is listed in Refs.[2, 3]. Earlier [1] we performed FEM with the coefficients $F_{ijkl}$ varied in a physically reasonable range, i.e., $|F_{ijkl}| \leq 10^{-11}$ m$^3$/C [4, 5], and assured that their influence is very small, and thus we neglect the role of $G_{flexo}$ hereinafter. The LGD parameters of a bulk ferroelectric SPS are listed in Ref.[1] and in **Table SI.**

**Table SI.** LGD parameters for a bulk ferroelectric Sn$_2$P$_2$S$_6$ (taken from Ref.[1])

| Parameter | Dimension | Parameters are taken from Refs. [6, 7, 8, 9, 10] |
|---|---|---|
| $\varepsilon_b$ | 1 | 41 |
| $\alpha_T$ | m/F | 1.44×10$^6$ |
| $T_C$ | K | 338 |
| $\beta$ | C$^{-4}$·m$^5$J | 9.40×10$^8$ |
| $\gamma$ | C$^{-6}$·m$^9$J | 5.11×10$^{10}$ |
| $g_{ij}$ | m$^3$/F | $g_{11}$ =5.0×10$^{-10}$, $g_{44}$=2.0×10$^{-10}$ |
| $s_{ij}$ | 1/Pa | $s_{11}$ =4.1×10$^{-12}$, $s_{12}$= −1.2×10$^{-12}$, $s_{44}$=5.0×10$^{-12}$ |
| $Q_{ij}$ | m$^4$/C$^2$ | Q$_{11}$=0.22, Q$_{12}$=0.12, Q$_{12}$ ≈ Q$_{13}$ ≈ Q$_{23}$ |

The relaxation-type Euler-Lagrange equation for $P_3$ has the form:

$$\Gamma \frac{\partial}{\partial t}P_3 + (\alpha - 2Q_{ij3}\sigma_{ij})P_3 + \beta P_3^3 + \gamma P_3^5 - g_{44}\left(\frac{\partial^2}{\partial x_1^2} + \frac{\partial^2}{\partial x_2^2}\right)P_3 - g_{11}\frac{\partial^2 P_3}{\partial x_3^2} = E_3. \tag{S.2}$$

The Khalatnikov kinetic coefficient is $\Gamma$ (see chapter 12 in Ref.[11]). The boundary condition for $P_3$ at the spherical surface is "natural", i.e., $\partial P_3/\partial r|_{r=R} = 0$. The potential $\phi$ satisfies the Poisson equation inside the particle,

$$\varepsilon_0\varepsilon_b\left(\frac{\partial^2}{\partial x_1^2} + \frac{\partial^2}{\partial x_2^2} + \frac{\partial^2}{\partial x_3^2}\right)\phi = -\frac{\partial P_3}{\partial x_3}, \tag{S.3a}$$

and the Laplace equation outside it,

$$\left(\frac{\partial^2}{\partial x_1^2} + \frac{\partial^2}{\partial x_2^2} + \frac{\partial^2}{\partial x_3^2}\right)\phi = 0. \tag{S.3b}$$

Equations (S.3) are supplemented by the condition of potential continuity at the particle surface, $(\phi_{ext} - \phi_{int})|_{r=R} = 0$. The boundary condition for the normal components of electric displacements is $\vec{n}(\vec{D}_{ext} - \vec{D}_{int})|_{r=R} = \sigma$, where the surface charge density $\sigma$ is given by Eq.(2) in the main text.

**Appendix B. Additional figures**

Dependences of the polarization $P_S$ on activity $a_i$ and $T$ calculated for the particle radius $R = 50$ nm and 100 nm are shown in **Fig. S1(a)-(d)**. Dependences of $P_S$ on $n_i$ and $R$ calculated for room temperature are shown in **Fig. S1(e)-(f)**.

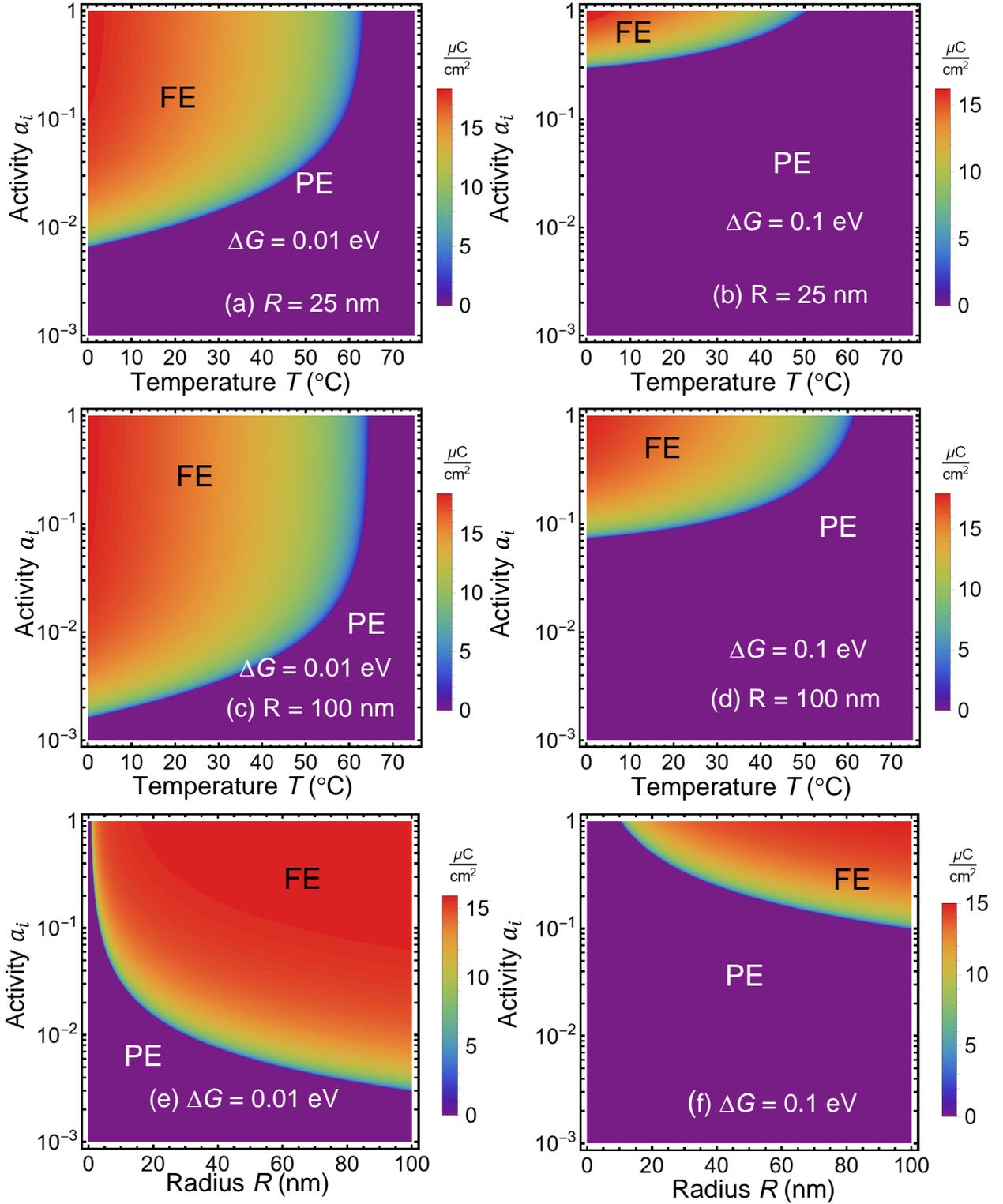

**Figure S1.** Dependences of the SPS nanoparticles spontaneous polarization $P_S$ on the surface charge activity $a_i$ and temperature $T$ calculated for the particle radius $R = 50$ nm (**a, b**) and 100 nm (**c, d**).

Dependences of $P_S$ on the activity $a_i$ and radius $R$ calculated for $T = 25^{\circ}$C **(e, f)**. The diagrams are calculated using the free energy (1) with LGD coefficients from **Table SI**, $Z_1 = -Z_2 = 1$, and $A_1 = A_2 \cong 0.16$ nm$^2$, $\varepsilon_b = 41$, $\varepsilon_e = 81$, $\Delta G_1 = \Delta G_2 \cong 0.01$ eV **(a, c, e)** and $\Delta G_1 = \Delta G_2 \cong 0.1$ eV **(b, d, f).**